\documentclass[summary]{emts2025}
\usepackage{cite}
\usepackage{amsmath,amssymb,amsfonts}
\usepackage{amsmath}
\usepackage{amsthm}

\usepackage{graphicx}
\usepackage{textcomp}
\usepackage{xcolor}
\usepackage{graphicx}
\usepackage{float}
\usepackage{subfigure}
\usepackage{amsmath}
\usepackage{amsfonts,amssymb}
\usepackage{mathrsfs}
\usepackage{mathtools}
\usepackage{algpseudocode}
\usepackage{bm}
\usepackage{multirow}
\usepackage{array}
\usepackage{amssymb}
\usepackage{amsmath}
\usepackage{cite}
\usepackage{url}
\usepackage{xcolor}
\usepackage{cite,graphicx,amsmath,amssymb}
\usepackage{subfigure}
\usepackage{fancyhdr}
\usepackage{mdwmath}
\usepackage{mdwtab}
\usepackage{caption}
\usepackage{amsthm}
\usepackage{setspace}
\usepackage{bm}
\usepackage{algorithm}
\usepackage{algpseudocode}
\usepackage{mathtools}
\usepackage{dsfont}
\usepackage{bbm}
\usepackage{multicol}
\usepackage{framed}
\newtheorem{remark}{Remark}
\newtheorem{theorem}{Theorem}

\newtheorem{lemma}{Lemma}

\newtheorem{corollary}{Corollary}

\makeatletter
\newcommand{\biggg}{\bBigg@{3}}
\newcommand{\Biggg}{\bBigg@{3.5}}
\makeatother

\title{Electromagnetic Channel Statistics for Continuous-Aperture Array (CAPA) Systems}

\author{Chongjun Ouyang\affref{ref1}, Boqun Zhao\affref{ref2}, Xingqi Zhang\affref{ref2},
  and Yuanwei Liu\affref{ref3}}

\affiliation{%
  \aff{ref1}{School of Electronic Engineering and Computer Science, Queen Mary University of London, London, E1 4NS, U.K.}
  \aff{ref2}{Department of Electrical and Computer Engineering, University of Alberta, Edmonton AB, T6G 2R3, Canada}
  \aff{ref3}{Department of Electrical and Electronic Engineering, The University of Hong Kong, Hong Kong}
}

\begin{document}

\maketitle

\begin{abstract}
The channel statistics of a continuous-aperture array (CAPA)-based channel are analyzed using its continuous electromagnetic (EM) properties. The received signal-to-noise ratio (SNR) is discussed under isotropic scattering conditions. Using Landau's theorem, the eigenvalues of the autocorrelation of the EM fading channel are shown to exhibit a step-like behavior. Building on this, closed-form expressions for the probability distribution of the SNR and the average channel capacity are derived. Numerical results are provided to validate the accuracy of the derivations.
\end{abstract}

\section{Introduction}
Multiple-antenna technology is a fundamental building block of modern wireless communication systems. In recent years, several novel array architectures have been proposed, such as massive multiple-input multiple-output (MIMO), holographic MIMO, and gigantic MIMO; see \cite{liu2024near} for more details. Despite their diverse architectures, these systems share a common evolutionary trend: larger aperture sizes, denser antenna configurations, higher operating frequencies, and more flexible structures \cite{ouyang2024diversity}. Naturally, this evolutionary trend leads to the creation of an (approximately) continuous electromagnetic (EM) aperture \cite{ouyang2024diversity}, i.e., a continuous-aperture array (CAPA).

A CAPA operates as a single large-aperture antenna with a continuous current distribution, comprising a (virtually) infinite number of radiating elements coupled with electronic circuits and fed by a limited number of radio-frequency (RF) chains \cite{ouyang2024diversity}. Compared with conventional discrete arrays, a CAPA fully utilizes the entire surface of the aperture and enables free control of the current distribution, thus yielding significantly improved performance. Against this background, CAPA has sparked increasing research interest; see \cite{liu2024near,ouyang2024diversity,ouyang2024primer} and references therein.

Due to its continuous nature, a CAPA-based channel should be modeled using EM theory and based on a continuous operator-based channel model, which differs significantly from the conventional matrix-based model for discrete arrays \cite{liu2024near,ouyang2024diversity,ouyang2024primer}. This makes it challenging to characterize the fundamental performance of CAPAs. It is further worth noting that existing research on CAPA-based channels is mostly focused on line-of-sight (LoS) channels, with the EM fading channel remaining less explored \cite{liu2024near}.

Motivated by the above knowledge gap, this paper aims to analyze the EM channel statistics of CAPA-based fading channels. Our goal is to characterize the statistics of the received signal-to-noise ratio (SNR) or the channel gain. The main contributions are summarized as follows: \romannumeral1) we analyze the received SNR of a CAPA-based channel considering isotropic fading; \romannumeral2) we use \emph{Landau's theorem} to characterize the eigenvalues of the autocorrelation of the EM fading channel; \romannumeral3) we further derive a closed-form expression for the statistical distribution of the received SNR, along with an expression for the channel capacity; \romannumeral4) we use numerical results to validate the effectiveness of the derived results and the superiority of CAPA over conventional MIMO in terms of channel capacity.

\begin{figure}[!t]
 \centering
\setlength{\abovecaptionskip}{0pt}
\includegraphics[height=0.15\textwidth]{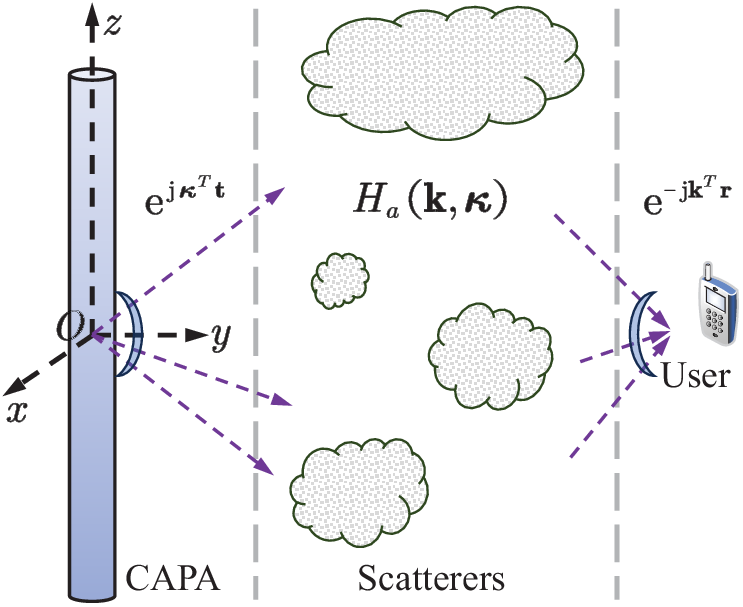}
\caption{Illustration of a CAPA-based channel.}
\label{Figure: System_Model}
\vspace{-15pt}
\end{figure}

\section{System Model}
\subsection{Signal Model}
We consider a point-to-point wireless communication system where the transmitter is equipped with a uni-polarized linear CAPA, as depicted in {\figurename} {\ref{Figure: System_Model}}. The CAPA is positioned along the $z$-axis and centered at the origin, with a physical length of $L$. The receiver is a single-antenna user. In a frequency-flat fading channel, the received signal at the user is expressed as follows \cite{pizzo2022spatial}: 
\begin{align}
y=\int_{{\mathcal{A}}}g({\mathbf{r}},{\mathbf{t}})x({\mathbf{t}})+n,
\end{align}
where ${\mathbf{r}}\in{\mathbbmss{R}}^{3\times1}$ denotes the location of the user, $n\sim{\mathcal{CN}}(0,\sigma^2)$ represents the complex Gaussian noise with $\sigma^2$ being the noise power, ${\mathcal{A}}=\{[0,0,z]^{T}|z\in[-\frac{L}{2},\frac{L}{2}]\}$ is the aperture of the CAPA, and $h({\mathbf{r}},{\mathbf{t}})$ characterizes the EM spatial response from ${\mathbf{t}}$ to ${\mathbf{r}}$. The transmit signal is given by
\begin{align}
x({\mathbf{t}})=\sqrt{P}j({\mathbf{t}})s,
\end{align}
where $s$ is the normalized coded data symbol satisfying ${\mathbbmss{E}}\{\lvert s\rvert^2\}=1$, $j({\mathbf{t}})$ denotes the normalized source current carrying the data information, subject to $\int_{{\mathcal{A}}}\lvert j({\mathbf{t}})\rvert^2{\rm{d}}{\mathbf{t}}=1$, and $P$ is the power budget. Consequently, the received signal-to-noise ratio (SNR) at the user is given by
\begin{align}
\gamma=\frac{P\left\lvert\int_{{\mathcal{A}}}h({\mathbf{r}},{\mathbf{t}})j({\mathbf{t}}){\rm{d}}{\mathbf{t}}\right\rvert^2}{\sigma^2}.
\end{align}
Note that the above signal model is integral-based, as opposed to matrix-based. To maximize the SNR, the source current should be aligned with the spatial response, which yields $j({\mathbf{t}})=\frac{h^{*}({\mathbf{r}},{\mathbf{t}})}{\sqrt{\int_{{\mathcal{A}}}\lvert h({\mathbf{r}},{\mathbf{t}})\rvert^2{\rm{d}}{\mathbf{t}}}}$ and, thus, $\gamma=\frac{P}{\sigma^2}\left\lvert\int_{{\mathcal{A}}}h({\mathbf{r}},{\mathbf{t}}){\rm{d}}{\mathbf{t}}\right\rvert^2$.
\subsection{Channel Model}
We assume that the scatterers are confined to the region between the CAPA and user. In this case, the EM multipath spatial response $h({\mathbf{r}},{\mathbf{t}})$ can be modeled as follows \cite{pizzo2022spatial,pizzo2020spatially}:
\begin{equation}\label{4FPWD_Model}
{h}({\mathbf{r}},{\mathbf{t}})=\iiiint_{{\mathcal{D}}({\bm\kappa})\times{\mathcal{D}}({\mathbf{k}})}
\frac{{\rm{e}}^{-{\rm{j}}{\mathbf{k}}^{T}{\mathbf{r}}}{H}({\mathbf{k}},{\bm\kappa})
 {\rm{e}}^{{\rm{j}}{\bm\kappa}^{T}{\mathbf{t}}}}{(2\pi)^2}{\rm{d}}{\mathbf{k}}{\rm{d}}{\bm\kappa},
\end{equation}
where ${\bm\kappa}=[\kappa_x,\gamma(\kappa_x,\kappa_z),\kappa_z]^{T}\in{\mathbbmss{R}}^{3\times1}$, ${\mathbf{k}}=[k_x,\gamma(k_x,k_z),k_z]^{T}\in{\mathbbmss{R}}^{3\times1}$, $\gamma(x,y)=(k_0^2-x^2-y^2)^{\frac{1}{2}}$, $k_0=\frac{2\pi}{\lambda}$ is the wavenumber, with $\lambda$ as the wavelength, ${\mathcal{D}}({\bm\kappa})=\{(\kappa_x,\kappa_y)\in{\mathbbmss{R}}^2|\kappa_x^2+\kappa_y^2\leq k_0^2\}$, and ${\mathcal{D}}({\mathbf{k}})=\{(k_x,k_y)\in{\mathbbmss{R}}^2|k_x^2+k_y^2\leq k_0^2\}$. Here, ${\rm{e}}^{{\rm{j}}{\bm\kappa}^{T}{\mathbf{t}}}$ and ${\rm{e}}^{-{\rm{j}}{\mathbf{k}}^{T}{\mathbf{r}}}$ denote the transmit and receive plane waves along the directions $\frac{\bm\kappa}{\lVert{\bm\kappa}\rVert}$ and $\frac{\mathbf{k}}{\lVert\mathbf{k}\rVert}$, respectively. Furthermore, $H({\mathbf{k}},{\bm\kappa})$ is the \emph{angular response} that maps each transmit direction $\frac{\bm\kappa}{\lVert{\bm\kappa}\rVert}$ to each receive direction $\frac{\mathbf{k}}{\lVert\mathbf{k}\rVert}$, which is modeled as a \emph{random process}. For analytical tractability, the isotropic scattering is considered, yielding \cite{pizzo2022spatial,pizzo2020spatially}
\begin{align}\label{Angular_Domain_Power_Distribution_Isotropic_Scattering}
H({\mathbf{k}},{\bm\kappa})=\frac{A_{s}(k_0)}{\sqrt{\gamma(k_x,k_z)\gamma(\kappa_x,\kappa_z)}}W({\mathbf{k}},{\bm\kappa}),
\end{align}
where $W({\mathbf{k}},{\bm\kappa})$ represents a \emph{zero-mean, unit-variance complex-Gaussian (ZUCG)} random field defined on ${\mathcal{D}}({\mathbf{k}})\times{\mathcal{D}}({\bm\kappa})$. Moreover, we set $A_{s}(k_0)=\frac{2\pi}{k_0}$ to normalize the channel power such that ${\mathbbmss{E}}\{\left\lvert h({\mathbf{r}},{\mathbf{t}})\right\rvert^2\}=1$. 
\section{EM Channel Statistics}
In the sequel, our aim is to characterize the statistics of this CAPA-based EM channel. Our focus will be on the statistical properties of the SNR.
\subsection{Statistical Equivalence}
Inserting \eqref{Angular_Domain_Power_Distribution_Isotropic_Scattering} into \eqref{4FPWD_Model} gives 
\begin{equation}\label{Spatial_Response_First}
{h}({\mathbf{r}},{\mathbf{t}})=\frac{1}{(2\pi)^2}\iint_{{\mathcal{D}}({\bm\kappa})}\frac{
 {\rm{e}}^{{\rm{j}}{\bm\kappa}^{T}{\mathbf{t}}}}{\sqrt{\gamma(\kappa_x,\kappa_z)}}\hat{H}({\bm\kappa}){\rm{d}}{\bm\kappa},
\end{equation}
where $\hat{H}({\bm\kappa})=\iint_{{\mathcal{D}}({\mathbf{k}})}
\frac{A_{s}(k_0){\rm{e}}^{-{\rm{j}}{\mathbf{k}}^{T}{\mathbf{r}}}}{\sqrt{\gamma(k_x,k_z)}}W({\mathbf{k}},{\bm\kappa}){\rm{d}}{\mathbf{k}}$. Since $W({\mathbf{k}},{\bm\kappa})$ is a complex Gaussian random field, $\hat{H}({\bm\kappa})$ is also a complex Gaussian random field. After some basic mathematical manipulations, it can be readily shown that $\hat{H}({\bm\kappa})\overset{d}{=}\hat{W}({\bm\kappa})A_{s}(k_0)(2\pi k_0)^{\frac{1}{2}}$, where $\overset{d}{=}$ denotes equivalence in distribution, and $\hat{W}({\bm\kappa})$ represents a \emph{ZUCG} random field defined on ${\mathcal{D}}({\mathbf{k}})$.

Noting that ${\bm\kappa}=[\kappa_x,\gamma(\kappa_x,\kappa_z),\kappa_z]^{T}$ and ${\mathbf{t}}=[0,0,z]^{T}$ for $z\in[-\frac{L}{2},\frac{L}{2}]$, we further simplify \eqref{Spatial_Response_First} as follows:
\begin{equation}\label{Spatial_Response_Second}
{h}({\mathbf{r}},{\mathbf{t}})=\frac{1}{(2\pi)^2}\int_{-k_0}^{k_0}{\rm{e}}^{{\rm{j}}\kappa_zz}H_z(\kappa_z){\rm{d}}\kappa_z\triangleq g(z),
\end{equation}
where $H_z(\kappa_z)=\int_{-\sqrt{k_0^2-\kappa_z^2}}^{\sqrt{k_0^2-\kappa_z^2}}\frac{
 \hat{H}({\bm\kappa})}{\sqrt{\gamma(\kappa_x,\kappa_z)}}{\rm{d}}\kappa_x$ is a complex Gaussian random field defined on $[-k_0,k_0]$. Since $\hat{H}({\bm\kappa})\overset{d}{=}\hat{W}({\bm\kappa})A_{s}(k_0)(2\pi k_0)^{\frac{1}{2}}$, it holds that 
 \begin{align}\label{Pre_Lemma_Autocorrelation_General_Result}
 H_z(\kappa_z)\overset{d}{=}A_{s}(k_0)(2\pi k_0)^{\frac{1}{2}}\pi^{\frac{1}{2}}W_z(\kappa_z),
 \end{align}
 where $W_z(\kappa_z)$ represents a \emph{ZUCG} random field defined on $[-k_0,k_0]$. The above arguments imply that $g(z)$ is a zero-mean complex Gaussian random field defined on $[-\frac{L}{2},\frac{L}{2}]$, whose statistics are determined by its autocorrelation function as follows:
 \begin{equation}\label{Lemma_Autocorrelation_General_Result}
 \begin{split}
R_{g}(z,z')&={\mathbbmss{E}}\{g(z)g^{*}(z')\}=
\frac{1}{(2\pi)^4}\int_{-k_0}^{k_0}\int_{-k_0}^{k_0}\\
&\times{\mathbbmss{E}}\{H_z(\kappa_z)H_z^{*}(\kappa_z')\}{\rm{e}}^{{\rm{j}}\kappa_zz}{\rm{e}}^{-{\rm{j}}\kappa_z'z'}{\rm{d}}\kappa_z
{\rm{d}}\kappa_z'.
\end{split}
\end{equation}
\subsection{Autocorrelation}\label{Section: Autocorrelation}
In this part, we analyze the properties of the autocorrelation function given in \eqref{Lemma_Autocorrelation_General_Result}. According to \eqref{Pre_Lemma_Autocorrelation_General_Result}, we have ${\mathbbmss{E}}\{H_z(\kappa_z)H_z^{*}(\kappa_z')\}=A_{s}^2(k_0)(2\pi k_0)\pi\delta(\kappa_z-\kappa_z')$, where $\delta(\cdot)$ denotes the Dirac delta function. It follows that 
\begin{align}
R_{g}(z,z')=\frac{1}{2k_0}\int_{-k_0}^{k_0}
{{\rm{e}}^{{\rm{j}}(z-z')\kappa_z}}{\rm{d}}\kappa_z,
\end{align}
which is a semipositive definite Hilbert–Schmidt operator. For clarity, we denote the eigendecomposition (EVD) of $R_{g}(z,z')$ as follows:
\begin{align}\label{EVD_Linear_Random_Operator}
R_{g}(z,z')=\sum\nolimits_{\ell=1}^{\infty}\sigma_{\ell}\phi_{\ell}(z)\phi_{\ell}^{*}(z'),
\end{align}
where $\sigma_{{\mathsf{r}}_x,1}\geq\sigma_{{\mathsf{r}}_x,2}\ldots\geq\sigma_{{\mathsf{r}}_x,\infty}\geq0$ are the eigenvalues of $R_{g}(z,z')$, and $\{\phi_{\ell}(\cdot)\}_{\ell=1}^{\infty}$ are the associated eigenfunctions that form an orthonormal basis over $[-\frac{L}{2},\frac{L}{2}]$. It follows that $\int_{-\frac{L}{2}}^{\frac{L}{2}}\phi_{\ell}^{*}(z)\phi_{\ell'}(z){\rm{d}}t_x=\delta_{\ell,\ell'}$, where $\delta_{\ell,\ell'}$ is the Kronecker delta. To glean further insights, we next unveil more properties of the eigenvalues. To this end, we define $K(z,z')\triangleq\frac{1}{2\pi}\int_{-k_0}^{k_0}{{\rm{e}}^{{\rm{j}}(z-z')\kappa_z}}{\rm{d}}\kappa_z$ for $z,z'\in[-\frac{L}{2},\frac{L}{2}]$, which yields $R_{g}(z,z')=\frac{2\pi}{2k_0}K(z,z')$. Let $\varepsilon_{1}\geq\varepsilon_{2}\ldots\geq\varepsilon_{\infty}\geq0$ denote the eigenvalues of $K(z,z')$. Then, $\frac{2\pi\varepsilon_{\ell}}{2k_0}=\sigma_{\ell}$. 

For an arbitrary square-integrable function $f(z')$ defined on $z'\in[-\frac{L}{2},\frac{L}{2}]$, we define $\hat{f}(z)\triangleq\int_{-\frac{L}{2}}^{\frac{L}{2}}K(z,z')f(z'){\rm{d}}z'$ for $z\in[-\frac{L}{2},\frac{L}{2}]$, which can be rewritten as follows:
\begin{align}
\hat{f}(z)={\mathbbmss{1}}_{[-\frac{L}{2},\frac{L}{2}]}(z)\int_{-\frac{L}{2}}^{\frac{L}{2}}k(z-z')
{\mathbbmss{1}}_{[-\frac{L}{2},\frac{L}{2}]}(z')f(z'){\rm{d}}z'.
\end{align}
The function $k(\cdot)$ is defined as $k(x)\triangleq \frac{1}{2\pi}\int_{-k_0}^{k_0}{{\rm{e}}^{{\rm{j}}x\kappa_z}}{\rm{d}}\kappa_z$ for $x\in{\mathbbmss{R}}$, which is the inverse Fourier transform of ${\mathbbmss{1}}_{[-k_0,k_0]}(\kappa_z)$. In other words, the Fourier transform of $k(x)$ ($x\in{\mathbbmss{R}}$) is an ideal filter over the range $[-k_0,k_0]$. Using this framework, the eigenvalues $\{\varepsilon_{\ell}\}_{\ell=1}^{\infty}$ can be characterized by \emph{Landau's eigenvalue theorem}, which states \cite{landau1975szego}:
\begin{align}\label{EDoF_Linear_Random_Operator_Statistical_Equal_Result1}
1\geq \varepsilon_{1}\geq\varepsilon_{2}\ldots\geq\varepsilon_{\infty}\geq0,
\end{align}
with $\{\varepsilon_{\ell}\}_{\ell=1}^{\infty}$ being functions of the number of effective degrees of freedom (DOFs):
\begin{align}
{\mathsf{DOF}}=\frac{1}{2\pi}\mu([-k_0,k_0])\mu([\begin{smallmatrix}-\frac{L}{2},\frac{L}{2}\end{smallmatrix}])=\frac{2k_0}{2\pi}L=\frac{2L}{\lambda}.
\end{align}
As $L\rightarrow\infty$ or ${\mathsf{DOF}}\rightarrow\infty$, the eigenvalues $\{\varepsilon_{\ell}\}_{\ell=1}^{\infty}$ \emph{polarize} asymptotically: for $\epsilon>0$, it holds that
\begin{equation} \label{EDoF_Linear_Random_Operator_Statistical_Equal_Result2}
\begin{split}
&\lvert\{\ell:\varepsilon_{\ell}>\epsilon\}\rvert={\mathsf{DOF}}\\
&+\left(\frac{1}{\pi^2}\log\frac{1-\sqrt{\epsilon}}{\sqrt{\epsilon}}\right)\log{{\mathsf{DOF}}}+o(\log{{\mathsf{DOF}}}),
\end{split}
\end{equation}
as $L\rightarrow\infty$ or ${\mathsf{DOF}}\rightarrow\infty$.
\vspace{-5pt}
\begin{remark}\label{remark_DOF}
The results in \eqref{EDoF_Linear_Random_Operator_Statistical_Equal_Result1} and \eqref{EDoF_Linear_Random_Operator_Statistical_Equal_Result2} imply that as $L$ increases, the leading ${\mathsf{DOF}}$ eigenvalues $\{\varepsilon_{\ell}\}_{\ell=1}^{{\mathsf{DOF}}}$ are near one, and the rest are near zero, with a transition band having a width proportional to $\log{{\mathsf{DOF}}}$. This reflects a behavior similar to that depicted in \cite[{\figurename} 2]{franceschetti2015landau}.
\end{remark}
\vspace{-5pt}
\subsection{SNR Distribution}
Since $g(z)$ is a zero-mean complex Gaussian random process, its statistics are entirely determined by its autocorrelation function $R_{g}(z,z')$. Based on \eqref{EVD_Linear_Random_Operator}, we obtain
\begin{align}\label{Linear_Random_Operator_Statistical_Equal_Result}
g(z)\overset{d}{=}{\overline{g}}(z)=\sum_{\ell=1}^{\infty}
\int_{-\frac{L}{2}}^{\frac{L}{2}}\sigma_{\ell}^{\frac{1}{2}}\phi_{\ell}(z)\phi_{\ell}^{*}(z'){\overline{W}}_z(z'){\rm{d}}z',
\end{align}
where ${\overline{W}}_z(z')$ denotes a \emph{ZUCG} random field over $[-\frac{L}{2},\frac{L}{2}]$. Note that \eqref{Linear_Random_Operator_Statistical_Equal_Result} holds because ${\overline{g}}(z)$ has the same autocorrelation as $g(z)$. It follows that the received SNR satisfies
\begin{align}\label{Linear_Random_Operator_Statistical_Equal_SNR}
\gamma=\overline{\gamma}\left\lvert\int_{{\mathcal{A}}}h({\mathbf{r}},{\mathbf{t}}){\rm{d}}{\mathbf{t}}\right\rvert^2
\overset{d}{=}\overline{\gamma}\int_{-\frac{L}{2}}^{\frac{L}{2}}\lvert {\overline{g}}(z)\rvert^2{\rm{d}}z,
\end{align}
where $\overline{\gamma}=\frac{P}{\sigma^2}$. Since ${\overline{W}}_z(z')$ is a ZUCG field, $\Phi_\ell=\int_{-\frac{L}{2}}^{\frac{L}{2}}\phi_{\ell}^{*}(z'){\overline{W}}_z(z'){\rm{d}}z'$ is a ZUCG random variable. Together with the orthogonality of $\{\phi_{\ell}(\cdot)\}_{\ell=1}^{\infty}$, this implies that $\{\Phi_{\ell}\}_{\ell=1}^{\infty}$ are independent and identically distributed (i.i.d.), with $\Phi_l\sim{\mathcal{CN}}(0,1)$ ($\forall \ell$). Therefore, $\gamma\overset{d}{=}\sum_{\ell=1}^{\infty}\sigma_{\ell}\lvert\Phi_{\ell}\rvert^2$.

Recalling Remark \ref{remark_DOF}, for small values of $\ell$, the eigenvalues $\epsilon_{\ell}$ or $\sigma_{\ell}$ decrease slowly until they reach a critical value at $\ell={\mathsf{DOF}}$, after which they decline rapidly. This step-like behavior becomes more pronounced as the physical length $L$ of the array increases. Since CAPAs are typically electromagnetically large arrays with $L\gg \lambda$, the sum $\sum_{\ell=1}^{\infty}\sigma_{\ell}\lvert\Phi_{\ell}\rvert^2$ is dominated by the first ${\mathsf{DOF}}$ terms, which yields
\begin{align}\label{EDoF_Linear_Random_Operator_Statistical_Equal_Result_Final}
\gamma\overset{d}{=}\sum\nolimits_{\ell=1}^{\infty}\sigma_{\ell}\lvert\Phi_{\ell}\rvert^2\approx\sum\nolimits_{\ell=1}^{{\mathsf{DOF}}}
\sigma_{\ell}\lvert\Phi_{\ell}\rvert^2.
\end{align}
In summary, the SNR can be approximated asymptotically as a finite weighted sum of exponentially distributed random variables. Based on this, the probability density function (PDF) of $\gamma$ is expressed as follows \cite{moschopoulos1985distribution}:
\begin{align}\label{PDF_Basic}
f_{\gamma}(x)=
\frac{\sigma_{\min}^{{\mathsf{DOF}}}}{\prod_{\ell=1}^{{\mathsf{DOF}}}\sigma_{\ell}}\sum_{q=0}^{\infty}\frac{\psi_q x^{{\mathsf{DOF}}+q-1}{\rm e}^{-\frac{x}{\sigma_{\min}}}}{\sigma_{\min}^{{\mathsf{DOF}}+q}\Gamma({\mathsf{DOF}}+q)},
\end{align}
where $\Gamma\left(z\right)=\int_{0}^{\infty}{\rm e}^{-t}t^{z-1}{\rm d}t$ is the Gamma function, $\sigma_{\min}=\min_{\ell\in\{1,\ldots,{\mathsf{DOF}}\}}\sigma_{\ell}=\sigma_{{\mathsf{DOF}}}$, and ${\psi _0} = 1$. The coefficients $\psi_q$ for $q\geq1$ are computed recursively as follows:
\begin{align}
{\psi _{q}} = \sum\nolimits_{k = 1}^{q} {\left[ {\sum\nolimits_{\ell = 1}^{{\mathsf{DOF}}}{{{\left( {1 - \sigma_{\min}/\sigma_{\ell}} \right)}^k}} } \right]} \frac{\psi _{q - k}}{{q}}.
\end{align}
\subsection{Performance Analysis}
Having characterized the statistics of the received SNR, we proceed to analyze the performance of the CAPA system. Our primary metric of interest is \emph{Shannon's channel capacity}, which is expressed as follows:
\begin{align}
C=\log_2(1+\gamma).
\end{align}
The average capacity is given by 
\begin{equation}\label{MISO_ADR_Explicit}
\begin{split}
{\mathbbmss{E}}\{C\}&=\frac{\sigma_{\min}^{{\mathsf{DOF}}}}{\prod_{\ell=1}^{{\mathsf{DOF}}}\sigma_{\ell}}
\sum_{q=0}^{\infty}\sum_{v=0}^{{\mathsf{DOF}}+q-1}
\frac{\psi_q/\log{2}}{({\mathsf{DOF}}+q-1-v)!}\\
&\times\left[\frac{(-1)^{{\mathsf{DOF}}+q-v}{\rm e}^{\frac{1}{{\overline{\gamma}}{\sigma_{\min}}}}}{({\overline{\gamma}}{\sigma_{\min}})^{{\mathsf{DOF}}+q-1-v}}{\rm{Ei}}
\left({\frac{-1}{{\overline{\gamma}}{\sigma_{\min}}}}\right)\right.\\
&+\left.\sum_{u=1}^{{\mathsf{DOF}}+q-1-v}\Gamma(u)\left(\frac{-1}{{\overline{\gamma}}{\sigma_{\min}}}\right)^{{\mathsf{DOF}}+q-1-v-u}\right],
\end{split}
\end{equation}
where ${\rm{Ei}}\left(x\right)=-\int_{-x}^{\infty}{\rm{e}}^{-t}t^{-1}{\rm{d}}t$ is the exponential integral function. The above expression is derived by using \eqref{PDF_Basic} and calculating the resultant integral. To derive further insights, we evaluate the capacity under the high-SNR regime by letting the transmit power tend to infinity:
\begin{align}\label{MISO_ADR_High_SNR_Asymptotic}
\lim\nolimits_{\overline{\gamma}\rightarrow\infty}{\mathbbmss{E}}\{C\}\simeq
{\mathcal{S}}(\log_2{\overline{\gamma}}-{\mathcal{L}}),
\end{align}
where ${\mathcal{S}}=1$, 
\begin{equation}
\begin{split}
{\mathcal{L}}=\frac{-\sigma_{\min}^{{\mathsf{DOF}}}}{\prod_{\ell=1}^{{\mathsf{DOF}}}\sigma_{\ell}}
\sum_{q=0}^{\infty}\frac{\psi_q
(\psi({\mathsf{DOF}}+q)+\log{\sigma_{\min}})}{\log{2}},
\end{split}
\end{equation}
and $\psi\left(x\right)=\frac{{\rm d}}{{\rm d}x}\ln{\Gamma\left(x\right)}$ is Euler's digamma function. These results indicate that the high-SNR slope, or the \emph{maximal multiplexing gain}, for the considered CAPA-based channel is ${\mathcal{S}}=1$, while ${\mathcal{L}}$ represents the corresponding high-SNR power offset in 3-dB units. Following similar arguments, other performance metrics such as the outage probability can also be derived. However, these details are omitted here due to page limitations.
\section{Numerical Results}
Numerical results are presented to validate the derived analytical findings. The following parameter setup is applied unless specified otherwise: the user is located at ${\mathbf{r}}=[500\lambda,0,0]^{T}$, the noise power is set as $\sigma^2=5.6\times10^{-3}~{\text{V}}^2/{\text{m}}$, and the carrier frequency is given by $f_{\rm{c}}=2.4$ GHz. The capacity achieved by CAPA is compared with conventional MIMO using a discrete uniform linear array with half-wavelength spacing, where each antenna element has a length of $\sqrt{{\lambda^2}/({4\pi})}$, akin to an isotropic antenna.

\begin{figure}[!t]
 \centering
\setlength{\abovecaptionskip}{0pt}
\includegraphics[height=0.21\textwidth]{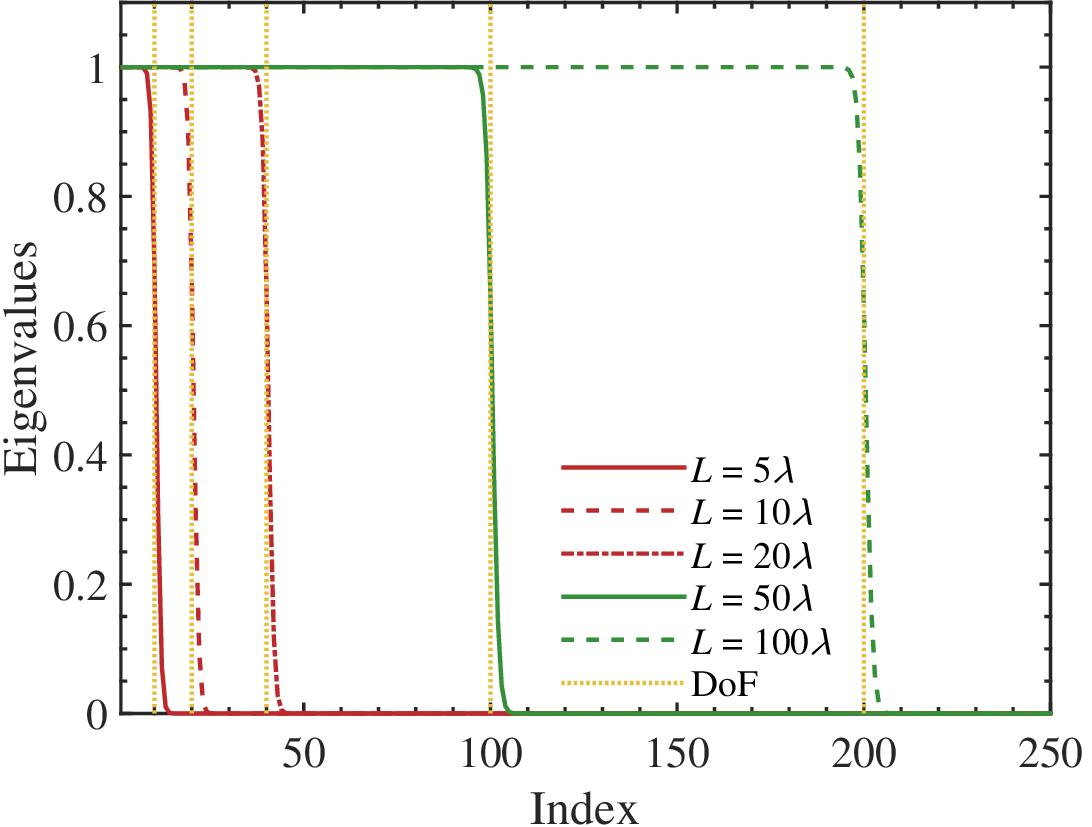}
\caption{Illustration of the eigenvalues of $K(z,z')$.}
\label{Figure: EVD}
\vspace{-10pt}
\end{figure}

{\figurename} {\ref{Figure: EVD}} illustrates the ordered eigenvalues of the operator $K(z,z')$ for different array lengths $L$, which are calculated using the method in \cite{atkinson2007solving}. As can be seen from this graph, the eigenvalues of $K(z,z')$ present a step-like behavior: exhibit a step-like behavior: when the index of the eigenvalue is smaller than ${\mathsf{DOF}}$, the eigenvalues decay slowly and are nearly one; however, when the index is larger than ${\mathsf{DOF}}$, the eigenvalues decay rapidly to zero. This validates the derivations in Section \ref{Section: Autocorrelation}. Additionally, we note that this step-like behavior is very pronounced even when $L$ is not very large, such as $L=10\lambda$. This establishes the foundation for our previous characterization of the received SNR by treating it as a finite weighted sum of exponentially distributed random variables.

\begin{figure}[!t]
 \centering
\setlength{\abovecaptionskip}{0pt}
\includegraphics[height=0.21\textwidth]{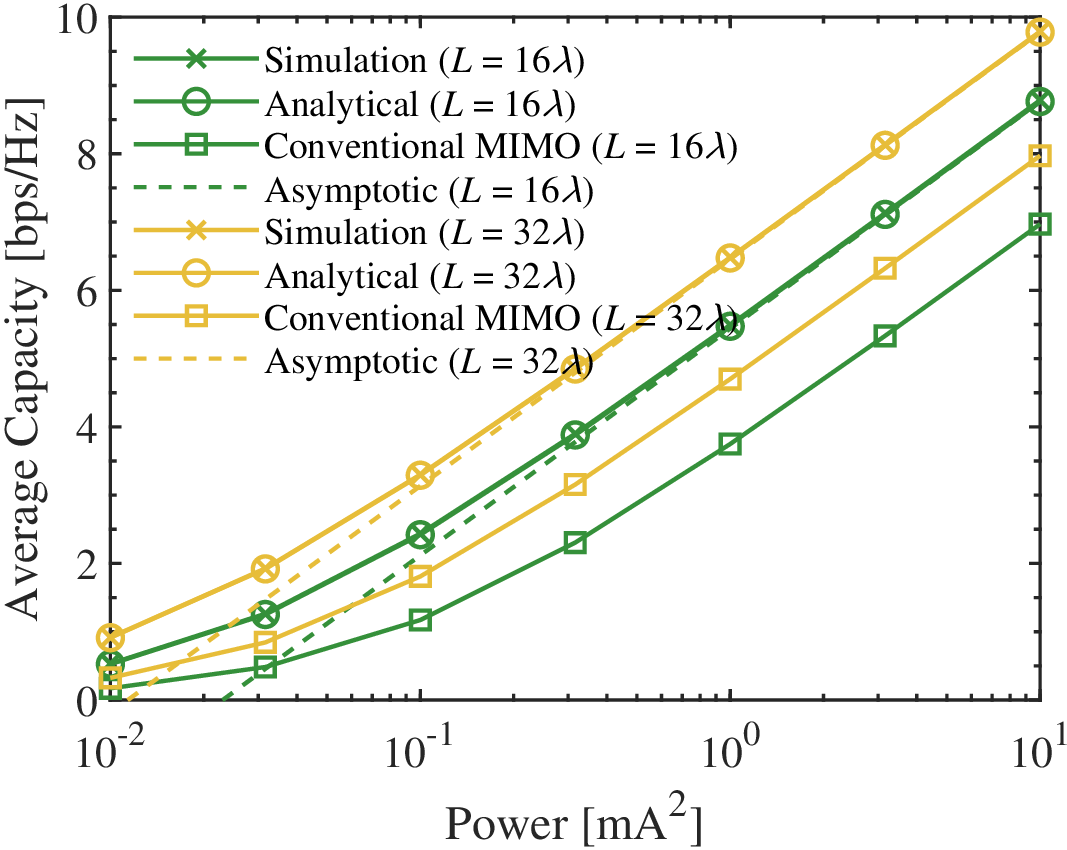}
\caption{Illustration of the average capacity.}
\label{Figure: Capacity}
\vspace{-15pt}
\end{figure}

{\figurename} {\ref{Figure: Capacity}} illustrates the average capacity, ${\mathbbmss{E}}\{C\}$, as a function of the transmit power $P$, where the analytical and asymptotic results are calculated using \eqref{MISO_ADR_Explicit} and \eqref{MISO_ADR_High_SNR_Asymptotic}, respectively. It can be observed that the analytical results match the simulations well, and the asymptotic capacity tracks closely with the analytical results in the high-SNR regime. This further validates the precision of the approximation in \eqref{EDoF_Linear_Random_Operator_Statistical_Equal_Result_Final} and our previous derivations. Moreover, we find that the average capacity achieved by CAPA is larger than that achieved by a conventional MIMO array. This superiority arises from the fact that CAPA makes full use of the spatial resources and benefits from greater spatial. 
\section{Conclusion}
We have derived novel expressions for the PDF of the SNR and the average capacity of CAPA-based fading channels. We have demonstrated that the received SNR can be statistically approximated by a finite weighted sum of several i.i.d. exponentially distributed variables with satisfying precision. We have also shown that CAPAs achieve a higher capacity than conventional MIMO arrays.

\end{document}